# Structured Ytterbium and Erbium -doped Silica Fiber for Dual Wavelength Laser Operation

Ivo Barton, Pavel Peterka, *Member, IEEE*, Martin Grabner *Member IEEE*, Jan Aubrecht, Michal Kamradek, Ondrej Podrazky, Petr Varak, Dariusz Pysz, Marcin Franczyk, Rafal Kasztelanic, Ryszard Buczynski, and Ivan Kasik

*Abstract*— We report on a novel type of dual-wavelength fiber laser with a structured-core design inside silica glass, forming a spatial separation of the several core areas doped with ytterbium and erbium ions. We have optimised the key parameters of the fiber core, such as the concentration of rare earth elements, and the optimal length of active fiber to operate simultaneously at two different wavelengths. Using the Modified Chemical Vapor Deposition method to obtain initial optical fiber preforms, and using the stack and draw technique, we have fabricated two types of active fibers, one with 7 and one with 19 rare-earth-doped rods (elements) forming the fiber core. We characterized the drawn fibers by investigating their structure by scanning electron microscopy, confirming the spatial separation of the elements within the core. Measuring absorption shows that concentration ratios $N_t^{Yb}$: $N_t^{Er}$ were approximately 52: 48 for 7 core fibers and 56:44 for 19 core fibers. Lifetimes for both active fibers were 0.84 ms for $Yb^{3+}$ ion and 10.30 ms for $Er^{3+}$ ion. The performance of fiber lasers was determined, proving that fibers are capable of laser emission simultaneously at 1042 nm and 1550 nm. We have shown experimentally that the output power ratio between both lasing wavelengths can be controlled by the length of the fiber.

## I. INTRODUCTION

Multi-wavelength fiber lasers are crucial for photonic component characterization, optical instrumentation, and optical sensing applications [1], [2], [3], [4]. In addition, narrow-linewidth, dual-wavelength lasers can generate microwave or mid-infrared radiation [5]. In the wavelength range of about 1000 nm – 1100 nm, $Yb^{3+}$-doped fiber lasers have shown excellent qualities, such as kW output powers [6], efficiencies exceeding 80 %, and very good $M^2$ values [7], [8]. At 1550 nm, fiber lasers are usually implemented using pigtailed semiconductor optical amplifiers or erbium-doped fiber amplifiers (EDFAs) as the gain medium [9], [10]. However, the nature of EDFA gain media usually prohibits multi-wavelength operation in EDFA lasers, especially for narrow wavelength spacings, unless techniques such as careful gain equalization are used [11]. Another dual-wavelength fiber laser was used to transmit millimeter waves for 5 G-supported radio over fiber links using specially designed erbium-doped fiber [12]. Fibers with $Er^{3+}/Yb^{3+}$ co-doped core (example of such fiber depicted in Fig. 1A) have been commonly used as fiber lasers or amplifiers. Such fibers can be excited by a pumping source operating at wavelengths of around 975 nm where $Yb^{3+}$ has its absorption peak, and because ions of $Yb^{3+}$ and $Er^{3+}$ are close to each other in a matrix, through non-radiative energy transfer between both of these, strong radiative emissions of $Er^{3+}$ ions is taking place [13], [14], [15]. The dual-wavelength fiber for the master oscillator power amplifier has been used to simultaneously amplify 1064 nm and 1548 nm signals in a niche application of different frequency generation in a periodically poled lithium niobate crystal for laser dispersion spectroscopy [16]. $Yb^{3+}/Er^{3+}$ doped fiber, prepared by the Modified Chemical Vapor Deposition (MCVD) method and polished fiber preform to obtain a D-shape [22], was employed for simultaneous emission at 1 μm and 1.5 μm. This approach achieved a pump power of 450 mW, with a power output of 42 mW at 1064 nm and 60 mW at 1550 nm. Dual-wavelength lasing was also achieved using a commercially available polarization-maintaining $Yb^{3+}/Er^{3+}$ co-doped fiber, which was modified by fabricating FBG pairs at 1057 nm and 1554 nm, respectively [17]. Dual emission was achieved with a power output of 0.65 W at 1057 nm and 0.64 W at 1554 nm. Such fiber has been used to pump a PPLN crystal and generate mid-infrared emission at 3.3 μm. Another example of the application of dual-wavelength fibers based on $Yb^{3+}/Er^{3+}$ fiber and hybrid $Yb^{3+}$ and $Yb^{3+}/Er^{3+}$ fibers has been engaged for low SWaP (Size, Weight, and Power) amplifiers for satellite communications at 1 μm and 1.55 μm [18]. It should be noted that in the case of $Yb^{3+}/Er^{3+}$ fibers designed for operation at 1.55 μm, dual operation occurs mainly at higher pump powers. In the co-doped $Yb^{3+}/Er^{3+}$ matrix, a situation arises where the $Er^{3+}$ ion's excited-state population cannot be further excited with increased pump power. This effect is known as the "bottleneck" effect. Excitation transfer rates between $Yb^{3+}$ and $Er^{3+}$ ions are strongly dependent on the distance between these ions, which, subsequently, depends on the concentrations [19]. The cross-

Ivo Barton, Pavel Peterka, Martin Grabner, Jan Aubrecht, Petr Varak, Michal Kamradek, Ondrej Podrazky, and Ivan Kasik are with the Institute of Photonics and Electronics of the Czech Academy of Sciences, Chaberská 1014, 182 00 Prague, Czechia. The corresponding author is Ivo Barton (e-mail: barton@ufe.cz). Marcin Franczyk, Dariusz Pysz, Rafal Kasztelanic, and Ryszard Buczynski are with Lukasiewicz Research Network, Institute of Microelectronics and Photonics, Al. Lotnikow 32/46, 02-668 Warsaw, Poland. Rafal Kasztelanic and Ryszard Buczynksi are also with Faculty of Physics University of Warsaw, Pasteura 5, Warsaw 02-093, Poland

This work was supported by the Czech Science Foundation (contract 21-45431L), Narodowe Centrum Nauki (OPUS LAP 020/33/IST7/02143) and co-funded by the EU project under the project LasApp CZ.02.01.01./00./22_008/0004573 relaxation process in $Yb^{3+}/Er^{3+}$ co-doped fiber is not fully controlled and may cause unstable laser operation or even fiber damage if pumped with high power. Thus, the $Yb^{3+}/Er^{3+}$ co-doped fiber lasers have not yet been realized as stable high-power dual-wavelength sources. Recently, $Yb^{3+}/Er^{3+}$ co-doped double-clad fiber was utilized in an amplifier setup with a dual-wavelength output power of up to 10 W, aiming to balance the output power by controlling the power of the seed signals [40].

The matrix accommodating rare earth (RE) ions is fundamental for generating dual-wavelength operations [20]. One of the possibilities was germanate glasses [21], where the matrix was co-doped with several RE ion species, i.e., $Yb^{3+}$, $Er^{3+}$, $Tm^{3+}$, and $Ho^{3+}$ [22], [23], [24]. Another matrix type was based on phosphate glasses because of high doping concentrations of trivalent rare-earth $(RE^{3+})$ ions exceeding 10 wt% [25], [26]. $Yb^{3+}$ and $Er^{3+}$ -doped phosphate glasses were employed to prepare a two-color laser operating simultaneously at 1040 nm and 1531 nm [27]. Silica glasses are another potential material for hosting REs in ultrafast fiber lasers [28], [20], [29], [30]. However, silica glass has several drawbacks, including high phonon energy and low RE ion solubility. To overcome these drawbacks, pure silica glass must be co-doped with a suitable modifier [31]. One such modifier is $P_2O_5$, with its unique high phonon energy (∼1200 cm−1), even higher than that of silica glass [32]. The high phonon energy makes $P_2O_5$ a suitable component for double-doped fibers, such as $Er^{3+}/Yb^{3+}$, because the high rates of multiphonon relaxation reduce the unwanted reverse energy transfer process from $Er^{3+}$ to $Yb^{3+}$[33]. $B_2O_3$ is commonly used as a network modifier in MCVD technology, but it exhibits a high phonon energy. Borosilicate glass systems are commonly used as a host matrix for the preparation of glass-ceramic fibers [34]. $GeO_2$ is another common modifier in optical fibers doped with RE [31]. Moderate concentrations of $GeO_2$ (<15 mol. %) are typically incorporated in optical fibers to increase the refractive index. Nevertheless, these low $GeO_2$ concentrations were shown to have a negligible influence on the emission properties of RE ions; the effect is much smaller than that of $Al_2O_3$ [35]. $Al_2O_3$ belongs amongst the most effective co-dopants of REs in silica glass [36], [37]. The $Al^{3+}$ ions prevent the formation of clusters and form a beneficial low-phonon solvation shell around the RE ions, thereby improving luminescence and laser properties [31]. Furthermore, another obstacle to preparing active fibers for dual-wavelength operation needs to be addressed: finding a suitable method that allows for the spatial distribution and separation of RE in the host matrix. Significant breakthroughs in nanoscience and nanotechnologies have recently opened new opportunities in optical fiber technology. The nanostructurization of fiber cores (example depicted in Fig. 1B) has emerged as an efficient method for preparing active fibers [38]. It offers the possibility to prepare large cores [39], otherwise impossible with standard techniques. The nanostructurization method, however, requires the manual assembly of a large number of elements, typically ranging from a few hundred to a few thousand, arranged into a lattice. The elements within the structure are very close to each other, which can be problematic in the case of a silica matrix doped with alumina, as at high temperatures of approximately 2000°C, diffusion of $Al_2O_3$ and RE can occur [40] potentially distorting the structure and properties of the designed fiber.

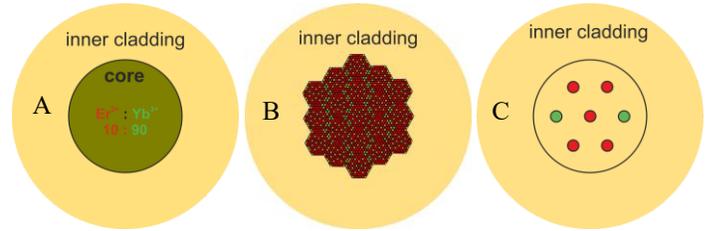

Fig. 1. Examples of different $Yb^{3+}$ and $Er^{3+}$ doping in fiber core in silica matrix A) Co-doped core with 90% $Yb^{3+}$+ 10% $Er^{3+}$.
B) Nanostructruralization of the core with hundreds of $Yb^{3+}$ and $Er^{3+}$ elements
C) Structuralized core with a limited number of $Yb^{3+}$ and $Er^{3+}$ elements

In this contribution, we present proof of concept of a novel type of active fiber laser (depicted in Fig. 1C) with a structured-core design inside silica glass that creates a spatial separation of the several areas doped with different REs in the single core. ;We used the stack and draw technique, similar to the method for nanostructured fibers [27]. However, in contrast to nanostructurization, we applied a limited number of RE rods (Fig. 1C) to mitigate the effects of diffusion and diminish interaction between $Yb^{3+}$ and $Er^{3+}$ ions in the fiber core structure. We have constructed a single active fiber, which can serve as the active medium at two different wavelengths corresponding to two REs inside one laser cavity while maintaining the output beam quality of a single-mode fiber. In contrast to the multi-10-W output power dual-wavelength, cladding-pumped $Yb^{3+}$-$Er^{3+}$ fiber lasers (that lase at two wavelengths thanks to the so-called bottleneck effect [41] at high output powers), we focus our work on proof-of-concept fiber that allows us to design core-pumped $Yb^{3+}$-$Er^{3+}$ fiber at almost arbitrary and stable output power, depending on the available pump power.

## II. DESIGN OF STRUCTURED FIBER

First, we briefly assess the design concerns of the cladding-pumped, dual-wavelength $Yb^{3+}/Er^{3+}$ fiber lasers. When using a commercially available $Yb^{3+}/Er^{3+}$ fiber that is intended for fiber lasers and amplifiers at 1.55 μm, the so-called "bottleneck" effect is used. However, that effect occurs only at higher powers, and controlled injection of seed signal at ~1 μm is often required for better control of laser operation [20]. The doping concentration in these fibers is optimized to maximize energy transfer from $Yb^{3+}$ to $Er^{3+}$ ions. Significantly improved control of the laser emission at both wavelengths, including lower power operation, was demonstrated using a cladding-pumped nanostructured fiber [27]. There, part of the core is singly $Yb^{3+}$ doped, providing a gain at ~1 μm, while the other part is $Yb^{3+}/Er^{3+}$ doped, thus providing a gain at ~1.55 μm, both with a single-wavelength single-pump laser. The doping concentration is designed to provide similar pump absorption at both areas, as the metastable level excitation, averaged along the length of the active fiber, is relatively low under cladding pumping. Since the fiber studied in this paper is core-pumped, it saturates the gain along a significant length of the active fiber, and different design criteria have to be used for finding $Yb^{3+}$ and $Er^{3+}$ concentrations that would allow similar power output at both 1 and 1.55 μm laser wavelengths.

Let us consider a Fabry-Perot fiber laser setup according to Fig. 2. The pigtail of the single-mode laser diode at 980 nm is spliced to concatenated highly reflective (>99 %) fiber Bragg gratings (FBGs) at 1042 nm and 1550 nm that form the wavelength-selective mirror of the laser. The two FBGs are spliced to the active fiber perpendicularly cleaved at the other end, thus serving as a Fresnel ~3.5 % laser output mirror. The FBGs are inscribed into a standard single-mode fiber G.652. Therefore, we assume to begin with the same waveguiding design of the active fiber, i.e., core radius $a$ of 4.15 μm and $NA$ of 0.13, and we search only for the optimal concentration and fiber lengths

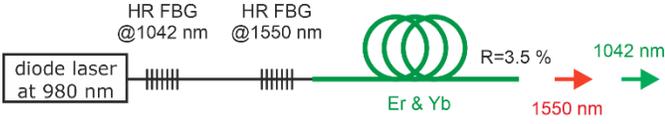

Fig. 2. Dual-wavelength fiber laser setup in core-pumped configuration.

For optimization of REs concentration, averaged across the fiber cross-section and the fiber length for given pump power, we used a numerical model of $Yb^{3+}/Er^{3+}$ fiber similar to commonly used numerical models based on simultaneous solution of laser rate equations and propagation equation. The principles of such numerical models are summarized, e.g., in [15], [42]. The energy level diagram of the $Yb^{3+}/Er^{3+}$ laser system is schematically shown in Fig. 3.

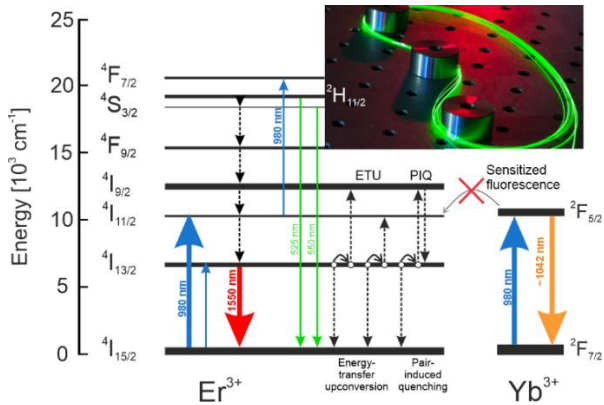

Fig. 3. Simplified energy level diagram of ytterbium erbium fiber with spatially separated erbium and ytterbium ions, where no energy transfer between Yb and Er is assumed. Dashed lines denote non-radiative processes. Inset: Optical fiber with core doped $Er^{3+}$ showing green emission due to excited state absorption.

Note that for the fiber under study, the two sets of ions, erbium and ytterbium ones, are separated in space, and hence, the energy transfer from $Yb^{3+}$ to $Er^{3+}$ ions is not considered in our case, contrary to the typical models of $Yb^{3+}/Er^{3+}$ fibers for high-power amplifiers at 1.55 μm. Note also that erbium fiber can shine green, as shown in the inset in Fig. 3, due to excited state absorption $^4I_{11/2} \rightarrow {}^2H_{11/2}, {}^3S_{3/2}$, thus forming a leakage channel that lowers the desired radiative transitions.

Based on the energy level diagram in Fig. 3, and neglecting upconversion processes to the levels above the $^4I_{11/2}$ level, one can write the laser rate equations as follows:

$$N_1^{Er} = N_{tot}^{Er} \frac{W_p^{Er} + W_a^{Er}}{W_p^{Er} + W_a^{Er} + W_e^{Er} - 1/\tau_{Er}} \qquad (1)$$

$$N_1^{Yb} = N_{tot}^{Yb} \frac{W_p^{Yb} + W_a^{Yb}}{W_p^{Yb} + W_a^{Yb} + W_e^{Yb} - 1/\tau_{Yb}} \qquad (2)$$

where $N_1^{Er}$, $N_{tot}^{Er}$, and $N_1^{Yb}$, $N_{tot}^{Yb}$ are the metastable level populations and total RE concentrations of the erbium and ytterbium, respectively; and $\tau_{Er}$, and $\tau_{Yb}$ are the fluorescence lifetimes of the erbium and ytterbium. Absorption, emission, and pump rates $W_a$, $W_e$, and $W_p$ of the erbium and ytterbium ions are given by:

$$W_{a,e}^{Er} = \int_0^\infty \sigma_{a,e}^{Er}(\lambda) \frac{P_\lambda(\lambda)\Gamma_\lambda}{h\nu\,\pi a^2}\,d\lambda, \qquad (3)$$

$$W_{a,e}^{Yb} = \int_0^\infty \sigma_{a,e}^{Yb}(\lambda) \frac{P_\lambda(\lambda)\Gamma_\lambda}{h\nu\,\pi a^2}\,d\lambda, \qquad (4)$$

$$W_p^{Er} = \frac{\Gamma_p \sigma_p^{Er} \mathbf{P_p^{980}}}{h\nu\,\pi a^2}, \qquad W_p^{Yb} = \frac{\Gamma_p \sigma_p^{Yb} \mathbf{P_p^{980}}}{h\nu\,\pi a^2} \qquad (5)$$

where $\sigma_a$ and $\sigma_e$ are the absorption and emission cross sections of the respective transition and rare earth element denoted in the indices, and $\Gamma$ is the overlap factor accounting for the overlap of the fundamental mode profile in polar coordinates $f_\lambda(r,\varphi)$ with the really doped area $N_r$:

$$\Gamma_\lambda = \iint_{r,\varphi} f_\lambda(r,\varphi)\,\frac{N_r(r,\varphi)}{N_t}\,r\,dr\,d\varphi \qquad (6)$$

where the integral of $N_r$ over the whole cross-section is equal to the product of the average total concentration $N_t$ and core area $\pi a^2$. We consider a generalized view of doping in the model, where both erbium and ytterbium ions are homogeneously distributed in the fiber core but do not interact with each other. Transforming the generalized view to the design of actual structured-core fibers, the model's total concentration of rare-earth ions corresponds to the average concentration of the respective rare-earth ions in the structured-core fiber, averaged over the fiber core area. On the one hand, the model is independent of the actual structure, but on the other hand, the calculated product of $\Gamma\,N_t$ may differ slightly between the generalized one and the real one given by eq. (6). Note that the only coupling of the erbium and ytterbium system in the fiber is through the typical pump at 980 nm, as apparent from formulae (5); as it is the key feature of the numerical model, $\mathbf{P_p^{980}}$ is highlighted by bold fonts.

Propagation equations describing the longitudinal evolution of all relevant light waves traveling in the positive direction of the z-axis are:

$$\frac{dP_\lambda^+}{dz} = \Gamma_\lambda P_\lambda^+ \left[ N_1^{Er}\sigma_{e\lambda}^{Er} - N_0^{Er}\sigma_{a\lambda}^{Er} - N_0^{Er}\sigma_{p\lambda}^{Er} \right.$$
$$\left. + N_1^{Yb}\sigma_{e\lambda}^{Yb} - N_0^{Yb}\sigma_{a\lambda}^{Yb} - N_0^{Yb}\sigma_{p\lambda}^{Yb} \right]$$
$$+ 2h\nu d\nu N_1^{Er}\sigma_{e\lambda}^{Er}$$
$$+ 2h\nu d\nu N_1^{Yb}\sigma_{e\lambda}^{Yb} - \chi_\lambda P_\lambda^+$$

Analogical equations can be written for the light waves travelling in the opposite direction. Note that the last three terms describe the spontaneous photons generated by erbium and ytterbium and the loss of the photons by the background attenuation $\chi_\lambda$, which should not be neglected in the structured core fibers, at least at the current state of the art. While we consider monochromatic radiation at the pump and laser wavelengths, the broad spectrum of ASE is split into partial waves in 1-nm wide wavelength bins. The propagation equations were numerically integrated by means of the Runge-Kutta method, and the iterative method had to be used because the boundary conditions at $z=L$ were not known at the beginning of the integration.

The numerical model was used to find the $N_t^{Yb}$ and $N_t^{Er}$ average concentration ratio, where the output power at both 1042 and 1550 nm is most similar/balanced for a given pump power. Note that if one considers identical local concentrations $N_t^{Yb}$ and $N_t^{Er}$, the task is equivalent to the search for the optimum $Yb^{3+}$:$Er^{3+}$ area ratio.

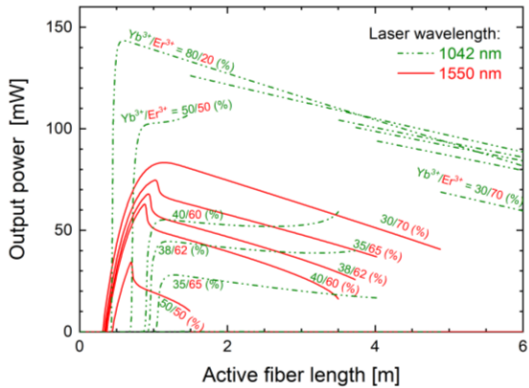

Fig. 4 Optimization of the average concentration of rare earth elements using numerical modeling. The share of ytterbium on the total RE concentration is noted next to each curve; the respective erbium concentration share is complement to 100 %. Pump power at 980 nm was set to 240 mW.

Figure 4 shows the fiber laser's calculated output powers at 1042 and 1550 nm for selected $Yb^{3+}$:$Er^{3+}$ average concentration ratios expressed as percentage of the number of $Yb^{3+}$ and $Er^{3+}$ ions. The pump power at 980 nm was set to 240 mW, i.e., the maximum coupled pump power available in the experiment described in the latter parts, and the background losses were set to 0.04 dB/m. Well-balanced output power at both wavelengths can be achieved for $Yb^{3+}$:$Er^{3+}$ with a ratio slightly below 40:60 %. For this ratio, the laser operates at a dual wavelength for a relatively broad range of fiber lengths, from about 1 m to 3.5 m long active fiber. For a higher $Yb^{3+}$ percentage, lasing at 1042 nm prevails, while for a higher $Er^{3+}$ share, it is the opposite. For a too high average $Yb^{3+}$ concentration, the lasing at 1550 nm does not occur at all, and analogically, for a too high average concentration of $Er^{3+}$, the lasing at 1042 nm is impossible, except for excessive length of the active fiber; see for example the case of $Yb^{3+}$ share of only 30 %, where ytterbium lasing at 1042 nm occurs at active fiber longer than about 5 m and dual wavelength operation does not occur at all. From Fig. 4, one can draw an approximate design constraint that the optimum average concentrations of ytterbium and erbium are similar for core-pumped dual-wavelength devices. It is in contrast with the cladding-pumped, dual-wavelength laser devices with low saturation, where the main design constraint is given by the similar pump absorption at both erbium-emission and ytterbium-emission areas.

### III. FIBER DEVELOPMENT

The preparation process of the $Yb^{3+}$ and $Er^{3+}$-dual-wavelength fiber laser (YbErDWFL) was done through several steps. The core of fiber has been assembled using the stack and draw technique. Two types of rods with different REs inside have been employed for assembly. The first type contained $Yb^{3+}$ ions, while the second contained $Er^{3+}$ ions. The preforms with REs were prepared using the MCVD and nanoparticle doping methods described in detail in [43], [44]. This method allows high-REs concentration without detrimental quenching effects. The first preform was composed of $Yb^{3+}$ in a silica-alumina matrix with a silica cladding. The second preform had the same silica-alumina matrix but with $Er^{3+}$.

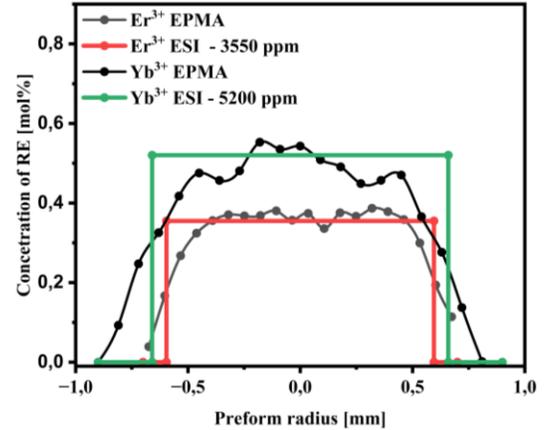

Fig. 5 Measured concentration by EPMA of MCVD preforms used for assembling of microstructured cores. Red ($Er^{3+}$) and green ($Yb^{3+}$) lines correspond to the equivalent step-index doping profile (ESI)

Fig. 5 shows measured concentration profiles obtained by wavelength-dispersive electron probe microanalysis (EPMA) using JEOL JXA-8230 and JXA-8800L analyzers. The dopant contents inside MCVD preforms were around 10 mol % of $Al_2O_3$, and the concentrations of RE corresponding to the Equivalent Step Index profile (ESI), see red and green lines in Fig 5, were 3550 mol ppm of $Er^{3+}$ and 5200 ppm for $Yb^{3+}$, respectively, to have an Al/RE ratio as high as possible. These preforms represent typical samples fabricated by the nanoparticle doping method [43], [44], in which the RE concentration was maximized while the laser device's efficiency is not severely affected by detrimental energy-transfer processes, e.g., pair-induced quenching. Note that the Al/REs ratio close to 50 was found desirable for high laser efficiency thanks to a low amount of unwanted energy transfer upconversion or clustering effects [44], [45]. The process of assembling the preform and, subsequently, the drawn fiber is depicted in Fig. 6. The process started with the doped preform being etched in concentrated fluoric acid to decrease the amount of cladding silica and change the ratio of the doped area and silica cladding, which also increases the average refractive index of the final fiber core. The preforms made by MCVD had inner/outer diameters of about 1.31/10 mm for $Yb^{3+}$ doped- and 1.35/9.8 mm for $Er^{3+}$ doped preform. The preform was 20 cm long, and etching was performed in a specially designed HF- resistant tube under rotation in the horizontal position to ensure longitudinal uniformity, with etching precision within 0.1 mm. After etching, the diameters ratio of the doped area to the undoped silica area was 1.31/2.5 mm for the $Yb^{3+}$-doped preform and 1.35/3 mm for the $Er^{3+}$-doped preform. This ratio was chosen to a) achieve the designed geometry (see Fig. 9) with suitable overlap factors Γ (see eq. (3)-(5)) determined from the mode fields and b) concurrently maintain the spatial separation of two different types of RE ions in the core to obtain dual-wavelength operation.

the initial concentration of RE in MCVD preforms. To achieve a 40/60 $Yb^{3+}/Er^{3+}$ concentration ratio for the hexagonal arrangement of a 7- rod core, two $Yb^{3+}$-doped rods and five $Er^{3+}$-doped rods were assembled within a silica capillary in a symmetric order, as shown in Fig. 6. It is necessary to note that the number and size of rods for each RE could be rearranged based on the initial concentration and size of MCVD preforms. The capillary with the rods was then overcladded with undoped silica tubes to form the final preform, visible in the lower part of the assembly process shown in Fig. 6. The final fiber was drawn at about 1940°C, resulting in a fiber with a core diameter of 7.2 μm and a silica cladding of 125 μm.

The overall macroscopic appearance of the fiber end face and detailed characterization of the fiber's core have been done using the SEM device TESCAN LYRA3 GM with a backscattered electron detector; the fiber sample was coated with a 20-nm-thick carbon layer.

The refractive index profile of the fabricated fiber was measured using a commercially available optical fiber analyzer (IFA-100, Interfiber Analysis Inc.) equipped with a laser source emitting at the wavelength of 979 nm with a spatial resolution of 500 nm. For fiber characterization, a 1-D scan and 2D tomography were used. In a 1-D scan, the interference fringes were first scanned passing the fiber and then passing the background without the fiber present. The data acquired from these 2 scans were analyzed to provide the axisymmetric refractive index profile. A one-dimensional (1-D) scan, shown in Fig. 8, was performed for a fiber laid in a fixed position by a single laser beam passing. In a 2-D tomography, data was acquired at 36 different angles of fiber rotation around its axis to provide a complete 2-D tomographic imaging of the fiber's internal refractive index profile.

The developed 7-rod-core, active fiber with micro-structured core guides light in a single transverse mode that does not generally match the mode shape of single-mode fiber used for pump and signal delivery. The splice losses between a single-mode fiber SMF28 and the active fiber were estimated by numerical computation of dominant modes propagating in both fibers. The modes were determined by a modal analysis using a finite element method in COMSOL software. Then, the modal overlap ratios Γ [-], as defined in equation (4) in [46], were computed from dominant mode field distributions of SMF28 and YbErDWFL fibers. The estimated splice losses $L$ [dB] were obtained as $L=-10 \log (G)$.

Background losses (attenuation) were measured using the standard cut-back method with a long fiber piece of 47 m and a short fiber piece of 2 m. A broadband halogen lamp, bare fiber adapters, and a monochromator-based spectrometer (Ando, AQ6317B with a spectral resolution of 2 nm) were used.

Fluorescence decay curves were measured using an EM4 P161-600-976 laser diode emitting at 976 nm. We used a Hamamatsu G8371-01 InGaAs PIN photodiode as a detector with a Thorlabs FEL1400 filter at wavelength of 1400 nm for the measurements of $Er^{3+}$ lifetime ($^4I_{13/2} \rightarrow {}^4I_{15/2}$ transition), and

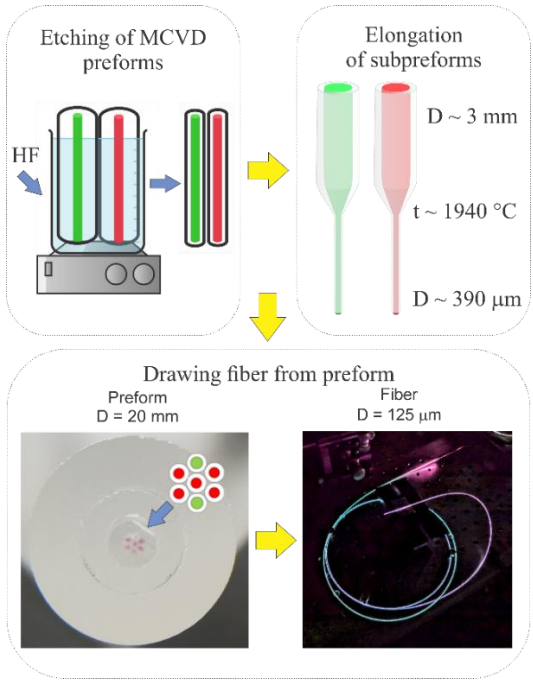

Fig. 6. The process of preparing the active fiber for the YbErDWFL,

The refractive index difference between the doped core and silica cladding was about 0.0245 for $Yb^{3+}$-doped preform and 0.021 for $Er^{3+}$-doped preform. Etched preforms were then elongated on the drawing tower at a temperature of 1930 °C into subpreforms (rods) with a diameter of approximately 390 μm. The number of rods for each RE in the assembly of the final preform was based on numerical modeling shown in Fig. 4 and

Thorlabs PDA36A Si photodiode detector with Thorlabs FEL1000 filter at wavelength 1000 nm for the measurements of $Yb^{3+}$ lifetime ($^2F_{5/2} \rightarrow ^2F_{7/2}$ transition). The measured fiber segment was approximately 5 cm long. The emission was detected from the side to suppress the influence of amplified spontaneous emission (ASE). The measurement setup and methodology are described in more detail in [47]. The decay curves were measured for multiple excitation powers, and a single-exponential fit of the measured data was used to obtain the lifetime values.

The concentration of OH groups in the spectral region around 1383 nm was measured using a SLED (Exalos, ESL1405-1111) source, and the transmitted intensities were detected with an Ando AQ6317B monochromator-based spectrometer over 1200-1600 nm, with a spectral resolution of 2 nm. To estimate the concentration of OH groups, tens of meters of a section (41 m for 7-core, and 20 m for 19-core) was spliced with passive single-mode fibers, and a modified cutback method was used.

The fibers were characterized by their spectral absorption. A tungsten halogen lamp was used as a broadband source of radiation. To determine the core absorption, we used a modified cutback method [48], where the transmission properties of the corresponding section of the active fiber, which was pigtailed from both sides by suitable passive fibers, and a reference fiber of the same type as used pigtails were measured and then compared. Absorption spectra were measured with a monochromator-based Ando AQ6317B spectrometer in the spectral range 600–1750 nm, with a resolution of 2 nm.

The performance of YbErDWFL was tested in Fabry-Perot configuration with a pump laser diode source operating at 974 nm (Lumics LU0975M450, Lumics GmbH) with a maximum output power of 450 mW, as can be seen in Fig. 2. The pump laser diode is equipped with an optical fiber isolator to protect the diode from accidental back reflections and the isolator losses are the main cause that the pump power coupled into the laser resonator is only 240 mW at maximum laser diode driver current. The laser cavities for both erbium and ytterbium lasers were formed by a single HRFBG (O/E land, reflecting at 1042 nm and 1550 nm, simultaneously), and the active fiber (YbErDWFL) that was perpendicularly cleaved at the output end to get a low-reflectivity mirror through Fresnel reflection. The optical filters (Thorlabs, FELH1000 or FELH1150) with absorption edges at 1 and 1.15 µm were gradually placed in a forward direction before the thermopile power detector (Gentec, XLP12-3S-H2-D0) to separate the pump and individual signal beams. The laser output spectra were recorded using an ANDO spectrometer at the highest pump power with a resolution of 0.5 nm. The particular laser output characteristics were calculated based on measurements without filters and then with the above-mentioned filters, with the knowledge of their transmission properties at corresponding wavelengths.

## IV. RESULTS

The appearance of the face of the drawn YbErDWFL can be observed in Fig.7. The fiber's core has a diameter of approximately 7.2 µm with a silica cladding of 125 µm. The core contains 7 separated bright spots representing $Yb^{3+}$- and $Er^{3+}$-doped areas. The core diameter was evaluated as the diameter of a circumscribed circle to all seven rods. A closer look at the fiber core through SEM can be observed in the inset in Fig. 7, which shows that $Er^{3+}$-doped areas (5 smaller ones) had a diameter of ~1.91 µm, and the $Yb^{3+}$-doped regions (2 bigger ones) had a diameter of ~2.16 µm. These doped areas are, however, larger than expected from an initial pure geometrical estimation. Theoretical areas were calculated from the dimensions of etched preforms and the draw-down ratio from preform to fiber. Estimated values were 1.29 µm for $Er^{3+}$-doped areas and 1.65 µm for $Yb^{3+}$-doped areas, showing an increase in size of 48 % for $Er^{3+}$-doped and 31% for $Yb^{3+}$-doped ones. The reason for enlarging doped areas can be explained by the diffusion processes at high temperatures during drawing and dwell time in the furnace [40], [49].

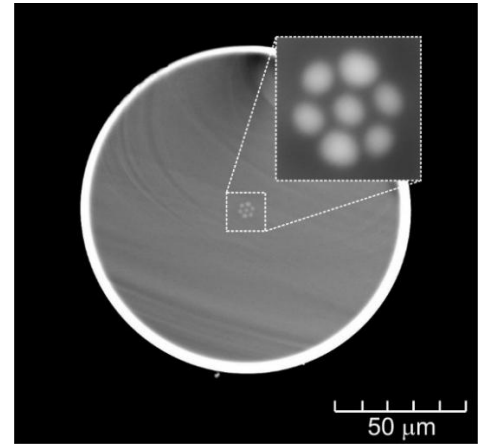

Fig. 7. SEM photo of whole fiber end face

The refractive index profile of YbErDWFL measured by IFA is depicted in Fig. 8. The 1D profile is averaged from five single scans around the fiber axis and shows three distinct peaks corresponding to three doped areas in one line. By analyzing the obtained results, we note that spaces between doped areas don't reach values of silica cladding, though dark spaces can be observed between doped areas in Fig. 7, suggesting the presence of undoped silica cladding. This can be explained by the spatial resolution of IFA of around 500 nm, which is close to 1 µm distance between the core areas. The refractive index difference between doped areas and cladding is at a level of 0.005, with a core diameter of about 7.8 µm, which agrees with values observed by electron microscopy. The refractive index difference has been used to calculate numerical aperture (NA), which for 978 nm corresponds to 0.12.

2D tomography of the refractive index profile of the fiber core is depicted in the top part of Fig. 8. The seven section of the core can be distinguished with two larger ones corresponding to $Yb^{3+}$-doped cores similar to results obtained by SEM. Smaller sections correspond to $Er^{3+}$-doped areas. We note that sections of the core seem connected; however, this could be explained in the same way as in the 1D scan by the limited spatial resolution of IFA.

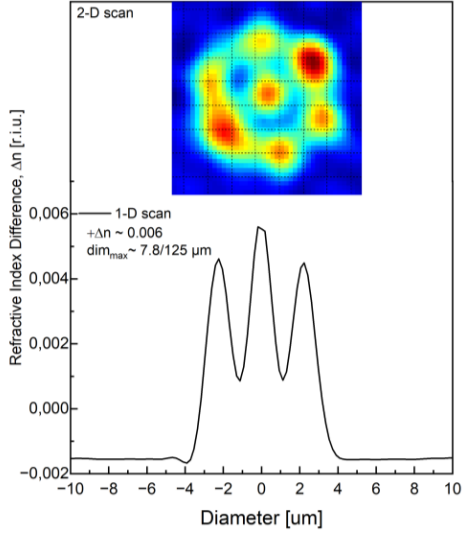

Fig. 8. Refractive index profile in 1D and 2D scans

The splice losses between SMF28 fiber and YbErDWFL were estimated by determining mode overlaps as described above. Fig. 9 shows the simulated fundamental mode for SMF28 fiber (left column) and YbErDWFL (right column). In both cases, the field is more squeezed to the fiber axis for a shorter wavelength. The calculated overlap factors are $\Gamma$=0.988 for λ=1042 nm and $\Gamma$=0.952 for λ =1550 nm. The splice losses are then $L$=-10log($\Gamma$), thus, about 0.05 dB for 1042 nm, respectively 0.21 dB for 1550 nm. The losses are not excessively large to prevent laser operation.

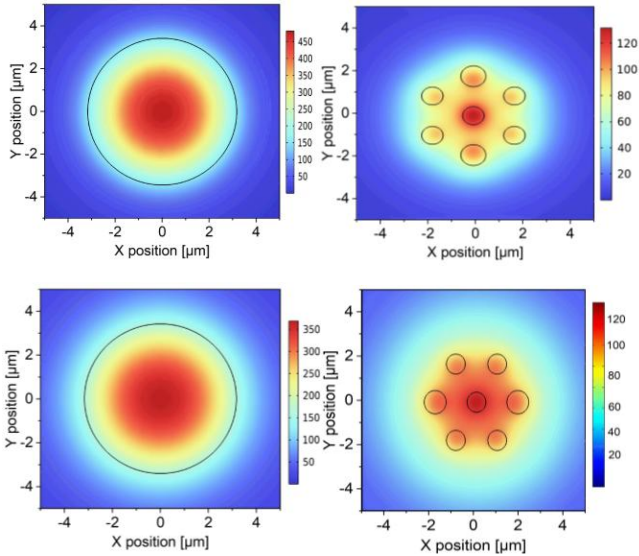

Fig 9. Fundamental mode electric field distribution from simulation of fibers SMF28 (left column) and YbErDWF (right column) for the wavelengths of A-B)1042 nm and C-D) 1550 nm.

.The fiber was characterized with regard to the fluorescence lifetime of $Er^{3+}$ and $Yb^{3+}$ ions. The measured fluorescence decay curves were single exponential, indicating distribution of both RE ions in a single optically active environment, the alumino silicate core matrix. The fluorescence lifetime of the $Er^{3+}$ ion in the $^4I_{13/2} \rightarrow ^4I_{15/2}$ transition was 10.39 ms, which is in good agreement with values typically found in erbium-doped silica fibers [31], [50], indicating no concentration quenching. The $Yb^{3+}$ ions exhibited a lifetime of 0.84 ms for the $^2F_{5/2} \rightarrow ^2F_{7/2}$. The lifetime of $Yb^{3+}$ ions in silica optical fibers is on average in range of 0.7 – 1 ms [51], [52]; the measured value fits in that interval. We have recently fabricated and investigated a $Yb^{3+}$-$Er^{3+}$ co-doped nanostructured-core fiber using identical preforms and a core composed of up to 10,000 elements, achieved by repeated stack-and-draw method [53]. The nanostructured-core fiber exhibited a nearly identical $Er^{3+}$ lifetime of 10.3 ms but a significantly shorter $Yb^{3+}$ lifetime of 0.76 ms; consequently, we were unable to achieve 1 μm laser emission. Therefore, the longer fluorescence lifetime of 0.84 ms indicates a potentially successful dual-wavelength operation. It has been demonstrated that the fluorescence lifetime of $Yb^{3+}$ can be shortened due to repeated heat treatment, likely due to diffusion and clustering of $Yb^{3+}$ ions, which result in harmful (for the intended purpose) energy-transfer processes between closely coupled ions [54],[55]. On the other hand, the lifetime of $Er^{3+}$ remains almost unchanged by the same effects [56]. Therefore, the fabrication of $Er^{3+}/Yb^{3+}$ co-doped silica fibers for dual-wavelength lasers requires careful optimization of the fabrication process, e.g., the number of heat treatment steps, to minimize the diffusion of $Yb^{3+}$ ions, prevent harmful ET processes, and achieve a long fluorescence lifetime of $Yb^{3+}$ ions.

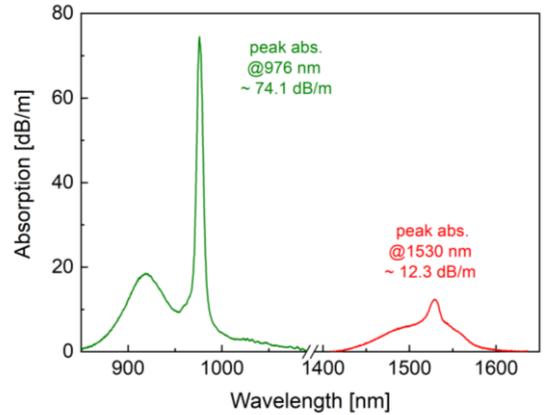

Fig 10. Absorption of 255-mm long fiber sample. Note that the absorption peak is composed of both erbium and ytterbium absorption.

The measured absorption spectrum of YbErDWFL is depicted in Fig. 10. For the YbErDWFL, absorption is expressed as:

$$\text{Abs}(\lambda) = 4{,}34 \times (\sigma_a^{Yb}\Gamma(\lambda) N_t^{Yb} + \sigma_a^{Er}\Gamma(\lambda) N_t^{Er}) \qquad (8)$$

While considering the absorption cross sections $\sigma_a^{Yb}$ and $\sigma_a^{Er}$ at the peak ground state absorption of ~ $25 \times 10^{-25}$ m$^2$ and ~ $2 \times 10^{-25}$ m$^2$ at 976 nm, respectively, and $\sigma_a^{Er}$ of $4.8 \times 10^{-25}$ m$^2$ at 1530 nm; and the overlap factor to be $\Gamma$=1 for simplicity, we got the $N_t^{Yb}$ fraction of 51.6 % of the total RE ions concentration.

The OH⁻ content was estimated as 3.1 mol ppm from the attenuation of 0.19 dB/m at the 1383 nm peak. In order to determine the OH- concentration, we used the attenuation 62.7 dB/(ppm × km) at the OH- peak in silica fibers [57].

Corresponding laser spectra measured at the output of the active fiber are displayed in Fig.11. Based on the previously mentioned simulation, the starting length of YbErDWFL was 7 m and gradually shortened to 1.5 m to find the optimum fiber length for the search for the optimum laser performance. As seen from Fig. 11, we found the optimum length of YbErDWFL to be 2 m, where the peak difference between ytterbium and erbium was about only 0.5 dB. This value agrees with the model presented in Fig. 4. The measured spectra confirm that the operating wavelength can be controlled by using different lengths of YbErDWFL fiber.

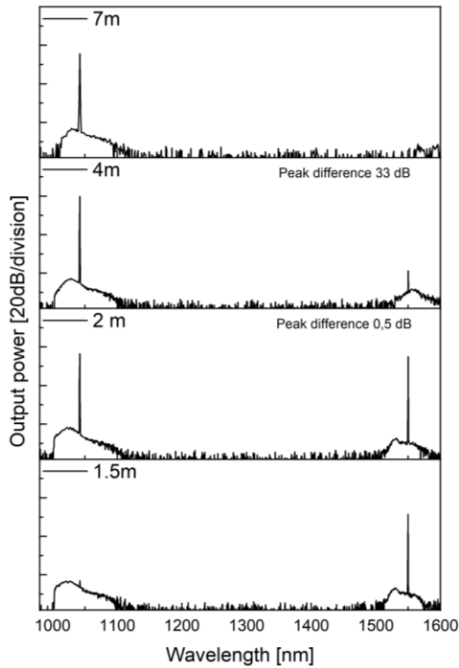

Fig. 11. Laser performance of YbErDFWL at different fiber lengths with pump power 220 mW

Laser curves, i.e., the dependence of laser output powers concerning pump (absorbed) powers for $Yb^{3+}$ and $Er^{3+}$ lasers, are shown in Fig. 12. The input pump power was measured immediately after the splice between the HRFBG and the active fiber to estimate the pump power launched correctly into the active fiber.

As can be seen from the laser curves as well as from recorded optical spectra, the maximum slope efficiency (SE) of 53.43% corresponding to $Yb^{3+}$ ions was achieved for a more extended fiber section of 4 m. Maximum output power was 72 mW. A significant laser peak corresponding to erbium ions at 1550 nm was observed only when the active fiber was gradually shortened (~ 3 m). The maximum SE corresponding $Er^{3+}$ was 23.89 % for 1.5m. The maximum output power reached for $Er^{3+}$ was 25 mW. The laser action at 1042 nm slowly decreases for short fiber sections and becomes negligible for a 1.5 m long active fiber. For the optimum fiber length of 2 m, the peak difference of only 0.5 dB was measured, with SE values of 27.7 % for $Yb^{3+}$ and 16.4 % for $Er^{3+}$ laser emission.

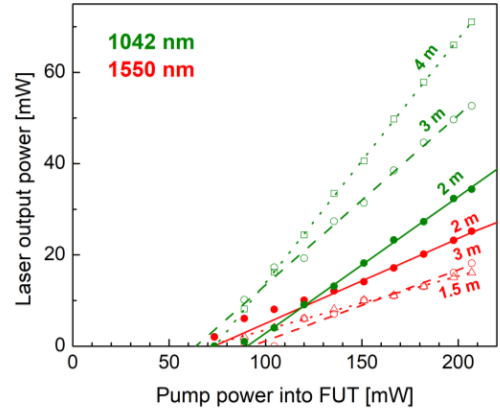

Fig. 12. Laser output power at 1042 and 1550 nm vs pump power for different lengths of active fiber.

**19-rod-core fiber**

After a successful demonstration of dual-wavelength laser operation with 7-element core fibers, one obvious question has arisen: Is it possible to achieve dual-wavelength emission with more elements in the core without the influence of diffusion from the drawing process? By using newly MCVD-made alumino-silicate preforms doped separately with $Yb^{3+}$ and $Er^{3+}$, we have assembled a new type of fiber with 19 doped rod elements inside the core. The core was assembled by 7 rods doped with $Yb^{3+}$ and 12 rods doped with $Er^{3+}$ ions. Doped rods were arranged in a hexagonal lattice, and $Yb^{3+}$ $Er^{3+}$ rods were evenly distributed.

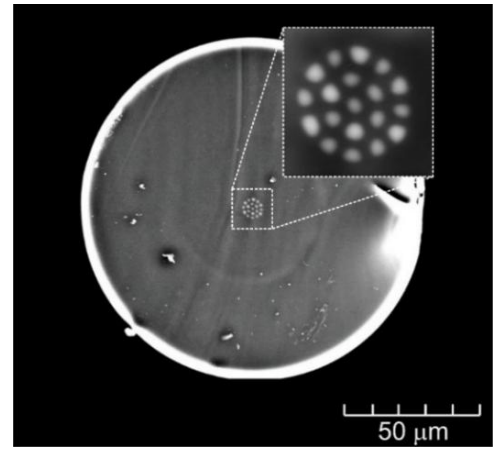

Fig. 13 SEM photo 19 core fiber

Fig. 13 shows the SEM photo of the appearance of the drawn 19-element core fiber (YbErDWFL-19). The fiber is composed of the core with a diameter of approximately 8.1 μm, with silica cladding with a diameter of about 124 μm. Bright spots, a sort of debris present on the face of fiber, are caused by cutting fiber and depositing 30 nm thick carbon layer. A detailed SEM close-up photo of the core is shown in the inset of Fig.13, where 19 clear, bright spots can be observed. Seven slightly brighter spots correspond to $Yb^{3+}$-doped rods with an average diameter of about 1.36 μm. Twelve smaller $Er^{3+}$-doped rods had an average diameter of about 1.22 μm. Again, we calculated the theoretical diameter for both types of doped rods from the dimensions of etched preforms and the draw-down ratio from preform to fiber. Theoretical diameters were 0.79 μm for $Yb^{3+}$ doped and 0.78 μm for $Er^{3+}$ doped ones. These values show an increase of rod diameters of 41 % for $Yb^{3+}$-doped and 36 % for $Er^{3+}$-doped. This agrees with 7-element core fiber, as the same diffusion occurs during high-temperature drawing.

IFA measured the refractive index profile of YbErDWFL-19, and the results corresponded with data obtained for 7-element core fiber. The refractive index difference between the doped core and cladding was at a level of 0.004, with a corresponding NA of 0.108. The fiber core diameter was about 8 μm. The $OH^-$ content for the 19-core fiber was estimated to be 0.72 mol ppm based on an attenuation of 0.045 dB/m at the 1383 nm peak. Like with the 7-core fiber, we determined the $OH^-$ concentration using the attenuation of 62.7 dB/(ppm × km) at the $OH^-$ peak in silica fibers [57].

The measured fluorescence lifetimes for 19-element core fiber were characterized in the same way as for 7-element core fiber with very similar results. The measured decay curves were again single exponential, suggesting a uniform distribution of both RE ions. The fluorescence lifetime of the $Er^{3+}$ ion in the $^4I_{13/2} \rightarrow ^4I_{15/2}$ transition was 10.20 ms, which agrees with the lifetime of 7-core fiber. The $Yb^{3+}$ ions had a lifetime of 0.84 ms, which is nearly identical with 7-elements core fiber. These values support the idea that a longer lifetime of $Yb^{3+}$ indicates successful dual-wavelength operation.

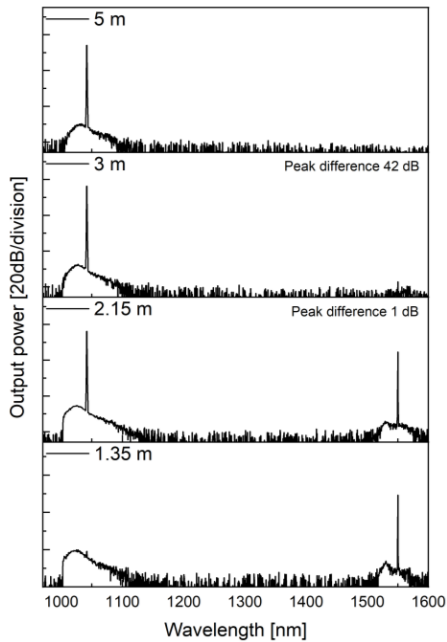

Fig.14 Laser performance of YbErDFWL with 19-element in core at different fiber lengths, with pump power 220 mW

We have used the same technique to characterize laser performance for 19-element core YbErDWFL with the same assumption and set-up as for 7-element core fiber. The $N_t^{Yb}$ fraction of 55.7 % of the total RE concentration was found from the peak absorption of 87.1 and 12.5 dB/m at 976 and 1530 nm, respectively.

The starting length of the 19-element core YbErDWFL was 5 m and gradually shortened to 1.3 m to find the optimum laser performance again. As seen from Fig. 14, we found the optimum length of 19-core YbErDWFL to be 2.15 m, where the peak difference between $Yb^{3+}$ and $Er^{3+}$ was about only 1 dB. This agrees with the results for the 7-core fiber and supports the presented model in Fig. 4.

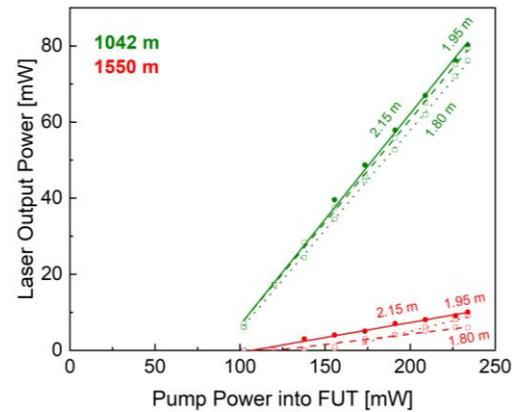

Fig 15. Laser output power at 1042 and 1550 nm vs pump power for the optimum length of 19-element in core active fiber

Laser curves describing the dependence of laser output power on pump (absorbed) powers for erbium and ytterbium lasers for 19 core fiber and optimum length of fiber are shown in Fig. 15. Measured values of SE were 55.98% at $Yb^{3+}$ and 11.5% at $Er^{3+}$ laser emission. Maximum output powers were 81.1 mW for $Yb^{3+}$ and 11 mW for $Er^{3+}$. A larger value of SE of $Yb^{3+}$ compared with SE of the same ions for optimal length of 7 core fiber can be explained by higher fraction of $Yb^{3+}$ ions in the total RE concentration in the 19-rod core fiber with respect to the 7-rod core fiber. Despite the difference of $Yb^{3+}$ ions fraction between the two fibers in only about 4 %, even such low difference may lead to significantly different behavior of the dual wavelength laser, as can be seen from Fig. 4.

## V. CONCLUSION

We have shown proof of concept of novel types of active fiber lasers in a silica matrix with the structured core operating simultaneously at two different wavelengths. We have optimized the key parameters of the fiber core. These parameters, such as the concentration of rare earth elements, the ratio between silica/REs, and the proper length of active fiber, could be employed to assemble active fiber. Using the MCVD and stack and draw techniques, we have fabricated two types of active fibers, one with 7 cores and one with 19 cores. We characterized our fibers by investigating the core structure with SEM, confirming the spatial separation of doped areas in the cores. Measuring absorption shows that concentration ratios $N_t^{Yb}$: $N_t^{Er}$ were approximately 52: 48 for 7-element core fibers and 56: 44 for 19-element core fibers. Lifetimes for both active fibers were on the level of 0.84 ms for $Yb^{3+}$ ion and 10.30 ms for $Er^{3+}$ ion. When measuring laser performance, we found the optimal length of 7-core fiber was 2 m and 2.15 m for fiber with 19 cores. We have shown experimentally that the length of the fiber can control the output power ratio between laser emission of $Yb^{3+}$ at 1042 nm and laser emission of $Er^{3+}$ at 1550 nm. Relatively low slope efficiencies and output power can be potentially increased by more precise control of the fiber laser length, precise control of the initial RE concentration in MCVD preforms. Output power and determined slope efficiencies for both wavelengths are dependent on the ratio of $Yb^{3+}$:$Er^{3+}$ as explained in Fig. 4. Further power scaling can be achieved in cladding-pumped arrangement. These structured fibers are intriguing and can be used to prepare special cores doped with rare earth elements like thulium or holmium for special active fibers.

## REFERENCES

[1] H. Wei, A. Amrithanath, and S. Krishnaswamy, "Multi-wavelength erbium-doped fiber ring lasers based on an optical fiber tip interferometer," in *Optical Components and Materials XVI*, M. J. Digonnet and S. Jiang, Eds., San Francisco, United States: SPIE, Feb. 2019, p. 78. doi: 10.1117/12.2511227.

[2] T. Li *et al.*, "Wavelength-switchable dual-wavelength thulium-doped fiber laser utilizing photonic crystal fiber," *Optics Communications*, vol. 528, p. 129033, Feb. 2023, doi: 10.1016/j.optcom.2022.129033.

[3] H.-Y. Pang, J.-T. Zheng, Z.-Y. Li, H. Liu, Z.-Y. Tao, and Y.-X. Fan, "A dual wavelength erbium-doped random fiber laser with tunable wavelength separation," *Optical Fiber Technology*, vol. 78, p. 103331, Jul. 2023, doi: 10.1016/j.yofte.2023.103331.

[4] R. A. Perez-Herrera *et al.*, "L-Band Multiwavelength Single-Longitudinal Mode Fiber Laser for Sensing Applications," *J. Lightwave Technol.*, vol. 30, no. 8, pp. 1173–1177, Apr. 2012.

[5] K. Krzempek, G. Sobon, J. Sotor, G. Dudzik, and K. M. Abramski, "Widely tunable, all-polarization maintaining, monolithic mid-infrared radiation source based on differential frequency generation in PPLN crystal," *Laser Physics Letters*, vol. 11, no. 10, p. 105103, Sep. 2014, doi: 10.1088/1612-2011/11/10/105103.

[6] W. Lin, M. Desjardins-Carrière, V. L. Iezzi, A. Vincelette, M.-H. Bussières-Hersir, and M. Rochette, "Simple design of Yb-doped fiber laser with an output power of 2 kW," *Optics & Laser Technology*, vol. 156, p. 108448, Dec. 2022, doi: 10.1016/j.optlastec.2022.108448.

[7] N. Platonov *et al.*, "High-efficient kW-level single-mode ytterbium fiber lasers in all-fiber format with diffraction-limited beam at wavelengths in 1000-1030 nm spectral range," in *Fiber Lasers XVII: Technology and Systems*, L. Dong and M. N. Zervas, Eds., San Francisco, United States: SPIE, Feb. 2020, p. 2. doi: 10.1117/12.2547021.

[8] M. Sohail *et al.*, "1.3 kW Continuous Wave Output Power of Ytterbium-Doped Large-Core Fiber Laser," *ECS J. Solid State Sci. Technol.*, vol. 10, no. 2, p. 026005, Feb. 2021, doi: 10.1149/2162-8777/abe6f6.

[9] A. Bellemare, M. Karasek, M. Rochette, S. LRochelle, and M. Tetu, "Room temperature multifrequency erbium-doped fiber lasers anchored on the ITU frequency grid," *Journal of Lightwave Technology*, vol. 18, no. 6, pp. 825–831, Jun. 2000, doi: 10.1109/50.848393.

[10] J. Cajzl *et al.*, "The influence of nanostructured optical fiber core matrix on the optical properties of EDFA," presented at the Proc.SPIE, May 2013, p. 877509. doi: 10.1117/12.2016922.

[11] M. Karásek, "The design of L-band EDFA for multiwavelength applications," *J. Opt. A: Pure Appl. Opt.*, vol. 3, no. 1, pp. 96–102, Jan. 2001, doi: 10.1088/1464-4258/3/1/316.

[12] H. J. Kbashi, V. Sharma, and S. Sergeyev, "Dual-wavelength fiber-laser-based transmission of millimeter waves for 5G-supported Radio-over-Fiber (RoF) links," *Optical Fiber Technology*, vol. 65, p. 102588, Sep. 2021, doi: 10.1016/j.yofte.2021.102588.

[13] W. L. Barnes, S. G. Crubb, and J. E. Townsend, "Yb3+ Sensitised Er3+ Doped Silica Optical Fibre with Ultra High Transfer Efficiency and Gain," *MRS Proceedings*, vol. 244, p. 143, 1991, doi: 10.1557/PROC-244-143.

[14] B. Dussardier, W. Blanc, and G. Monnom, "Luminescent Ions in Silica-Based Optical Fibers," *Fiber and Integrated Optics*, vol. 27, no. 6, pp. 484–504, Nov. 2008, doi: 10.1080/01468030802269746.

[15] P. Peterka and J. Vojtěch, "Optical Amplification," in *Handbook of Radio and Optical Networks Convergence*, T. Kawanishi, Ed., Singapore: Springer Nature Singapore, 2023, pp. 1–51. doi: 10.1007/978-981-33-4999-5_20-1.

[16] K. Krzempek, G. Sobon, J. Sotor, and K. M. Abramski, "Fully-integrated dual-wavelength all-fiber source for mode-locked square-shaped mid-IR pulse generation via DFG in PPLN," *Opt. Express*, vol. 23, no. 25, p. 32080, Dec. 2015, doi: 10.1364/OE.23.032080.

[17] A. R. EL-Damak, Jianhua Chang, Jian Sun, Changqing Xu, and Xijia Gu, "Dual-Wavelength, Linearly Polarized All-Fiber Laser With High Extinction Ratio," *IEEE Photonics J.*, vol. 5, no. 4, pp. 1501406–1501406, Aug. 2013, doi: 10.1109/JPHOT.2013.2276991.

[18] P. Steinvurzel, W. W. Chang, N. I. Werner, and T. S. Rose, "Dual band optical amplifier for 1.0 and 1.5 μm communications," in *Free-Space Laser Communications XXXIV*, H. Hemmati and B. S. Robinson, Eds., San Francisco, United States: SPIE, Mar. 2022, p. 14. doi: 10.1117/12.2608082.

[19] D. K. Vysokikh, A. P. Bazakutsa, A. V. Dorofeenko, and O. V. Butov, "Experiment-based model of an Er/Yb gain medium for fiber amplifiers and lasers," *J. Opt. Soc. Am. B*, vol. 40, no. 9, p. 2273, Sep. 2023, doi: 10.1364/JOSAB.497167.

[20] P. D. Dragic, M. Cavillon, and J. Ballato, "Materials for optical fiber lasers: A review," *Applied Physics Reviews*, vol. 5, no. 4, p. 041301, Oct. 2018, doi: 10.1063/1.5048410.

[21] F. B. Slimen *et al.*, "Highly efficient \-Tm3+ doped germanate large mode area single mode fiber laser," *Opt. Mater. Express*, vol. 9, no. 10, pp. 4115–4125, Oct. 2019, doi: 10.1364/OME.9.004115.

[22] Z. Yang, S. Xu, L. Hu, and Z. Jiang, "Thermal analysis and optical properties of Yb3+/Er3+-codoped oxyfluoride germanate glasses," *J. Opt. Soc. Am. B*, vol. 21, no. 5, pp. 951–957, May 2004, doi: 10.1364/JOSAB.21.000951.

[23] R. Wang, Z. Yang, D. Zhou, Z. Song, and J. Qiu, "Structure and luminescent property of Er3+-doped germanate glasses," *Journal of Non-Crystalline Solids*, vol. 383, pp. 200–204, Jan. 2014, doi: 10.1016/j.jnoncrysol.2013.02.032.

[24] M. Khalid, D. G. Lancaster, and H. Ebendorff-Heidepriem, "Spectroscopic analysis and laser simulations of Yb3+;/Ho3+; co-doped lead-germanate glass," *Opt. Mater. Express*, vol. 10, no. 11, pp. 2819–2833, Nov. 2020, doi: 10.1364/OME.404375.

[25] A. Adel, M. Farag, M. El-Okr, T. Elrasasi, and M. El-Mansy, "Preparation and Characterization of Phosphate Glasses Co-doped with Rare Earth Ions," *Egypt. J. Chem.*, vol. 0, no. 0, pp. 0–0, Oct. 2019, doi: 10.21608/ejchem.2019.15556.1944.

[26] W. C. Wang, B. Zhou, S. H. Xu, Z. M. Yang, and Q. Y. Zhang, "Recent advances in soft optical glass fiber and fiber lasers," *Progress in Materials Science*, vol. 101, pp. 90–171, Apr. 2019, doi: 10.1016/j.pmatsci.2018.11.003.

[27] M. Franczyk *et al.*, "Dual Band Active Nanostructured Core Fiber for Two-Color Fiber Laser Operation," *Journal of Lightwave Technology*, vol. 40, no. 21, pp. 7180–7190, 2022, doi: 10.1109/JLT.2022.3199581.

[28] F. Todorov *et al.*, "Active Optical Fibers and Components for Fiber Lasers Emitting in the 2-μm Spectral Range," *Materials*, vol. 13, no. 22, p. 5177, Nov. 2020, doi: 10.3390/ma13225177.

[29] G. E. Keiser, "A Review of WDM Technology and Applications," *Optical Fiber Technology*, vol. 5, no. 1, pp. 3–39, Jan. 1999, doi: 10.1006/ofte.1998.0275.

[30] P. F. Kashaykin *et al.*, "Radiation induced attenuation in pure silica polarization maintaining fibers," *Journal of Non-Crystalline Solids*, vol. 508, pp. 26–32, Mar. 2019, doi: 10.1016/j.jnoncrysol.2018.12.016.

[31] P. Vařák *et al.*, "Luminescence and laser properties of RE-doped silica optical fibers: The role of composition, fabrication processing, and inter-ionic energy transfers," *Optical Materials: X*, vol. 15, p. 100177, Aug. 2022, doi: 10.1016/j.omx.2022.100177.

[32] H. Ebendorff-Heidepriem and D. Ehrt, "Spectroscopic properties of Eu3+ and Tb3+ ions for local structure investigations of fluoride

phosphate and phosphate glasses," *Journal of Non-Crystalline Solids*, vol. 208, no. 3, pp. 205–216, Dec. 1996, doi: 10.1016/S0022-3093(96)00524-8.

[33] Y. Kobayashi *et al.*, "Effect of P-to-Rare Earth Atomic Ratio on Energy Transfer in Er-Yb-Doped Optical Fiber," *J. Lightwave Technol.*, vol. 38, no. 16, pp. 4504–4512, Aug. 2020, doi: 10.1109/JLT.2020.2989774.

[34] S. Kang *et al.*, "Precisely controllable fabrication of $Er^{3+}$-doped glass ceramic fibers: novel mid-infrared fiber laser materials," *J. Mater. Chem. C*, vol. 5, no. 18, pp. 4549–4556, 2017, doi: 10.1039/C7TC00988G.

[35] W. Blanc *et al.*, "Thulium environment in a silica doped optical fibre," *Journal of Non-Crystalline Solids*, vol. 354, no. 2–9, pp. 435–439, Jan. 2008, doi: 10.1016/j.jnoncrysol.2007.06.083.

[36] O. Podrazky, I. Kasik, M. Pospisilova, and V. Matejec, "Use of alumina nanoparticles for preparation of erbium-doped fibers," in *LEOS 2007 - IEEE Lasers and Electro-Optics Society Annual Meeting Conference Proceedings*, Lake Buena Vista, FL, USA: IEEE, Oct. 2007, pp. 246–247. doi: 10.1109/LEOS.2007.4382369.

[37] C. C. Baker *et al.*, "Nanoparticle doping for high power fiber lasers at eye-safer wavelengths," *Opt. Express*, vol. 25, no. 12, pp. 13903–13915, Jun. 2017, doi: 10.1364/OE.25.013903.

[38] M. Franczyk, R. Stępień, A. Filipkowski, D. Pysz, and R. Buczyński, "Nanostructured Core Active Fiber Based on Ytterbium Doped Phosphate Glass," *J. Lightwave Technol.*, vol. 37, no. 23, pp. 5885–5891, Dec. 2019.

[39] M. Franczyk, K. Stawicki, J. Lisowska, D. Michalik, A. Filipkowski, and R. Buczyński, "Numerical Studies on Large-Mode Area Fibers With Nanostructured Core for Fiber Lasers," *J. Lightwave Technol.*, vol. 36, no. 23, pp. 5334–5343, Dec. 2018.

[40] J. Kirchhof, S. Unger, and B. Knappe, "Diffusion Processes in Lightguide Materials," in *Optical Fiber Communication Conference*, Optica Publishing Group, 2000, p. WM1. [Online]. Available: https://opg.optica.org/abstract.cfm?URI=OFC-2000-WM1

[41] Z. Chen, X. Cheng, X. Zeng, H. Jiang, X. Yang, and Y. Feng, "Balancing dual-band output in Er/Yb co-doped fiber amplifier," *Optical Fiber Technology*, vol. 87, p. 103930, Oct. 2024, doi: 10.1016/j.yofte.2024.103930.

[42] M. Karasek, "Optimum design of Er/sup 3+/-Yb/sup 3+/ codoped fibers for large-signal high-pump-power applications," *IEEE Journal of Quantum Electronics*, vol. 33, no. 10, pp. 1699–1705, Oct. 1997, doi: 10.1109/3.631268.

[43] O. Podrazky, I. Kasik, M. Pospíšilová, and V. Matejec, *Use of alumina nanoparticles for preparation of erbium-doped fibers*. 2007, p. 247. doi: 10.1109/LEOS.2007.4382369.

[44] M. Kamrádek *et al.*, "Nanoparticle and Solution Doping for Efficient Holmium Fiber Lasers," *IEEE Photonics Journal*, vol. 11, no. 5, pp. 1–10, 2019, doi: 10.1109/JPHOT.2019.2940747.

[45] P. Vařák *et al.*, "Nanoparticle doping and molten-core methods towards highly thulium-doped silica fibers for 0.79 μm-pumped 2 μm fiber lasers – A fluorescence lifetime study," *Journal of Luminescence*, vol. 275, p. 120835, 2024, doi: https://doi.org/10.1016/j.jlumin.2024.120835.

[46] B. Bai, L. Liu, R. Chen, and Z. Zhou, "Low Loss, Compact TM-Pass Polarizer Based on Hybrid Plasmonic Grating," *IEEE Photonics Technology Letters*, vol. 29, no. 7, pp. 607–610, 2017, doi: 10.1109/LPT.2017.2663439.

[47] P. Vařák *et al.*, "Heat treatment and fiber drawing effect on the luminescence properties of RE-doped optical fibers (RE-Yb, Tm, Ho)," *Opt. Express*, vol. 30, no. 6, pp. 10050–10062, Mar. 2022, doi: 10.1364/OE.449643.

[48] B. Jiřičková, M. Grábner, C. Jauregui, J. Aubrecht, O. Schreiber, and P. Peterka, "Temperature-dependent cross section spectra for thulium-doped fiber lasers," *Opt. Lett.*, vol. 48, no. 3, pp. 811–814, Feb. 2023, doi: 10.1364/OL.479313.

[49] J. Kirchhof, S. Unger, and A. Schwuchow, "Fiber lasers: materials, structures and technologies," in *Optical Fibers and Sensors for Medical Applications III*, I. Gannot, Ed., SPIE, 2003, pp. 1–15. doi: 10.1117/12.498062.

[50] M. Kamrádek *et al.*, "Energy transfer coefficients in thulium-doped silica fibers," *Opt. Mater. Express*, vol. 11, no. 6, pp. 1805–1814, Jun. 2021, doi: 10.1364/OME.427456.

[51] A. Schwuchow, S. Unger, S. Jetschke, and J. Kirchhof, "Advanced attenuation and fluorescence measurement methods in the investigation of photodarkening and related properties of ytterbium-doped fibers," *Appl. Opt.*, vol. 53, no. 7, pp. 1466–1473, Mar. 2014, doi: 10.1364/AO.53.001466.

[52] R. Paschotta, J. Nilsson, P. R. Barber, J. E. Caplen, A. C. Tropper, and D. C. Hanna, "Lifetime quenching in Yb-doped fibres," *Optics Communications*, vol. 136, no. 5, pp. 375–378, Apr. 1997, doi: 10.1016/S0030-4018(96)00720-1.

[53] I. Barton *et al.*, "Optimization of erbium and ytterbium concentration in nanostructured core fiber for dual-wavelength fiber lasers," in *Specialty Optical Fibres*, K. Kalli, A. Mendez, and P. Peterka, Eds., SPIE, 2023, p. 1257311. doi: 10.1117/12.2666703.

[54] P. Vařák *et al.*, "Heat treatment and fiber drawing effect on the luminescence properties of RE-doped optical fibers (RE=Yb, Tm, Ho)," *Opt. Express*, vol. 30, no. 6, pp. 10050–10062, Mar. 2022, doi: 10.1364/OE.449643.

[55] E. A. Savel'ev, A. V. Krivovichev, and K. M. Golant, "Clustering of Yb in silica-based glasses synthesized by SPCVD," *Optical Materials*, vol. 62, pp. 518–526, Dec. 2016, doi: 10.1016/j.optmat.2016.11.004.

[56] P. Vařák *et al.*, "Heat treatment and fiber drawing effect on the matrix structure and fluorescence lifetime of Er- and Tm-doped silica optical fibers," *Opt. Mater. Express*, vol. 14, no. 4, p. 1048, Apr. 2024, doi: 10.1364/OME.520422.

[57] J. Aubrecht *et al.*, "Self-swept holmium fiber laser near 2100 nm," *Opt. Express*, vol. 25, no. 4, p. 4120, Feb. 2017, doi: 10.1364/OE.25.004120.

Authors

**Disclosure: The authors declare no conflicts of interest**

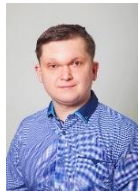

Ivo Barton received an M.Sc. degree in 2012 from the University of Chemistry and Technology, where he developed optical fibers for the photocatalytic decomposition of organic compounds, for which the Czech Glass Society awarded him for the best diploma thesis. He received a Ph.D. degree in 2019 from the University of Chemistry and Technology for preparing new types of optical fibers for delivering high energies. He is currently working as an Associate scientist, developing novel types of optical fibers for fiber lasers.

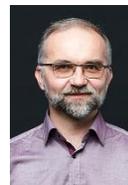

Pavel Peterka (Member, IEEE) received the M.Sc. degree in physical engineering and the Ph.D. degree in radioelectronics from Czech Technical University, Prague, Czech Republic, in 1993 and 2000, respectively. From 2001 to 2003, he worked for 13 months with Laboratoire de Physique de la Matière Condensée (LPMC), CNRS - Université de Nice – Sophia Antipolis, Nice, France, on thulium–doped fibers. He is currently the Director of the Institute of Photonics and Electronics, Czech Academy of Sciences, Prague. His research interests include numerical modeling and spectroscopic characterization of rare-earth doped fibers, development of specialty fibers and components for fiber lasers. As an Associate Professor of applied physics, he is also active as a teacher of a semestral course on fiber lasers and amplifiers with Czech Technical University.

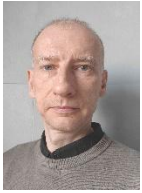

Martin Grábner received his M.Sc. and Ph.D. degrees both in radio electronics from the Czech Technical University in Prague in 2000 and 2008, respectively. He has been working in the area of electromagnetic wave propagation in an inhomogeneous and random media. In the group of fiber lasers, his current research interests include numerical modeling of light propagation in active and passive optical waveguides.

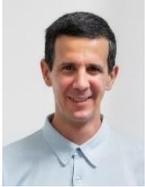

Jan Aubrecht received his M.Sc. degree from the Czech Technical University in Prague in 2006. In his master's thesis, he focused on the development of a fiber-optic sensor for gaseous ammonia. In 2014 he received his PhD degree at the Czech Technical University in Prague for the development of fiber-optic sensors. Currently, he works as a research scientist on the development of characterization methods of active optical fibers and on the development of fiber lasers and amplifiers.

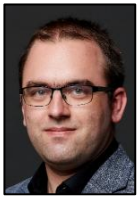

Petr Varak received his Ph.D. degree in Inorganic Chemistry and Material Engineering from the University of Chemistry and Technology, Prague, for developing novel silicate glass and glass-ceramics materials for photonic applications. He is currently working as a post-doctoral fellow at the Institute of Photonics and Electronics, as well as an assistant professor at the University of Chemistry and Technology, Prague. His main areas of focus are the synthesis of silicate glass and glass-ceramics materials for optics and photonics, and their structural and spectroscopic characterization.

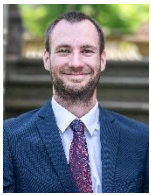

Michal Kamrádek received his M.Sc. degree in chemistry from the University of Chemistry and Technology, Prague, Czechia in 2012. He received a Ph.D. degree from the Czech Technical University in Prague, Faculty of Nuclear Sciences and Physical Engineering in 2022. Since 2016, he has been working at the Institute of Photonics and Electronics, Czech Academy of Sciences. His research interests include optical fiber technology and fiber lasers.

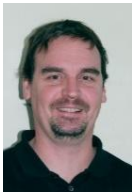

Ondrej Podrazky received his PhD degree in environmental chemistry and technology from the Institute of Chemical Technology Prague in collaboration with the Czech Academy of Sciences in 2003. Between 1999 and 2005, he dealt with optical biosensors at the Institute of Chemical Process Fundamentals, Czech Academy of Sciences, Prague. Since 2006, he has worked as a research fellow and since 2024 as a senior research fellow at the Institute of Photonics and Electronics, Czech Academy of Sciences, Prague. He focuses on the development of technology for the fabrication of specialty optical fibers for fiber lasers, fiber-based sensors, and bioresorbable fiber applications.

Rafal Kasztelanic - Lukasiewicz Research Network Institute of Microelectronics and Photonics, Warsaw, Poland, Faculty of Physics, University of Warsaw, Warsaw, Poland

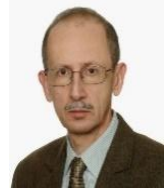

Dariusz Pysz received the Eng. and M.Sc. degrees in technical physics (optics) from the Warsaw University of Technology, Warsaw, Poland, in 1985. In 1986, he joined the Department of Glass, Institute of Electronic Materials Technology, Warsaw. He has more than 20 years of hands-on experience in glass fiber drawing, including silica and soft-glass step index fibers, and more recently, photonic crystal fibers and hollow core fibers. His primary research interests include fiber optics, image-guided structures, and photonic crystal fiber technology.

Marcin Franczyk received the M.Sc. degree from the Warsaw University of Technology, Warsaw, Poland, in 1999, and the Ph.D. degree in materials science from the Institute of Electronic Materials Technology, Warsaw, Poland, in 2012. In 1998 and 1999, he gained his optoelectronics experience from the Centre for Photonics and Photonic Materials, Bath, U.K., working on the manufacturing and characterization of photonic crystal fibers. He is currently an Assistant Professor with the Institute of Microelectronics and Photonics, Warsaw. His current research interests include fiber lasers, large-mode area fibers, and microoptics.

Ryszard Buczynski received the Ph.D. degree in physics from the Warsaw University of Technology, Warsaw, Poland, in 1999, and the second Ph.D. degree in applied science from Vrije Universiteit Brussel, Brussels, Belgium, in 2000. He was a Postdoctoral Fellow with Vrije Universiteit Brussel and with Heriot-Watt University, Edinburgh, U.K. He is currently a Professor with the Institute of Microelectronics and Photonics, Warsaw, and with the Faculty of Physics, University of Warsaw, Warsaw. He leads an inter-university research team Microoptics and Photonic Crystal Fibers Group. He is a co-author of more than 100 publications in peer-reviewed journals. His research interests include photonic crystal fibers, supercontinuum generation, microoptics, and microfluidics.

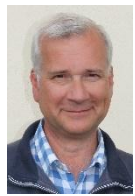

Ivan Kasik received the Ph.D. degree in technology of silicates from the Institute of Chemical Technology, Prague, Czech Republic, in 1995, for research on silica optical fibers. Since then, he is a Senior Research Fellow of the Institute of Photonics and Electronics, Academy of Sciences of the Czech Republic, Prague. He deals with materials for the preparation of optical fibers, methods of preparation, and fiber characterization. He is currently a Co-Principal Investigator of the project Novel nanostructured optical fibers for fiber lasers operating at dual wavelengths, related to this paper. He is a member of SPIE, Czech, and Slovak Society of Photonics and Czech Glass Society.